\documentclass[prc,showpacs,superscriptaddress,twocolumn]{revtex4}

\usepackage{amsmath,bm}
\usepackage{graphicx}
\usepackage{epstopdf}
\usepackage{subfigure}
\usepackage{epsfig}
\usepackage{amsmath,amssymb,amsfonts}
\usepackage[normalem]{ulem}
\usepackage{color}
\usepackage[utf8]{inputenc}
\usepackage{verbatim}
\usepackage{array}
\usepackage{mathtools}
\usepackage{booktabs, ctable}
\usepackage{xcolor, color, framed}
\usepackage[colorlinks, linkcolor=red,anchorcolor=green,citecolor=blue]{hyperref}

\begin{document}

\newcommand{\vk}{{\vec k}}
\newcommand{\vK}{{\vec K}}
\newcommand{\vb}{{\vec b}}
\newcommand{{\vp}}{{\vec p}}
\newcommand{{\vq}}{{\vec q}}
\newcommand{\vQ}{{\vec Q}}
\newcommand{\vx}{{\vec x}}
\newcommand{\dd}{\textrm{\,d}}
\newcommand{\beq}{\begin{equation}}
\newcommand{\eeq}{\end{equation}}
\newcommand{\half}{{\textstyle \frac{1}{2}}}
\newcommand{\gton}{\stackrel{>}{\sim}}
\newcommand{\lton}{\mathrel{\lower.9ex \hbox{$\stackrel{\displaystyle<}{\sim}$}}}
\newcommand{\ee}{\end{equation}}
\newcommand{\ben}{\begin{enumerate}}
\newcommand{\een}{\end{enumerate}}
\newcommand{\bit}{\begin{itemize}}
\newcommand{\eit}{\end{itemize}}
\newcommand{\bc}{\begin{center}}
\newcommand{\ec}{\end{center}}
\newcommand{\bea}{\begin{eqnarray}}
\newcommand{\eea}{\end{eqnarray}}

\newcommand{\beqar}{\begin{eqnarray}}
\newcommand{\eeqar}[1]{\label{#1} \end{eqnarray}}
\newcommand{\pleft}{\stackrel{\leftarrow}{\partial}}
\newcommand{\pright}{\stackrel{\rightarrow}{\partial}}

\newcommand{\eq}[1]{Eq.~(\ref{#1})}
\newcommand{\fig}[1]{Fig.~\ref{#1}}
\newcommand{\eff}{ef\!f}
\newcommand{\alphas}{\alpha_s}

\renewcommand{\topfraction}{0.85}
\renewcommand{\textfraction}{0.1}

\renewcommand{\floatpagefraction}{0.75}
\newcommand{\pp}{p\kern-0.05em p}
\newcommand{\ppbar}{\mathrm{p}\kern-0.05em \overline{\mathrm{p}}}
\newcommand{\pPb}{\ensuremath{\mbox{p--Pb}}}
\newcommand{\PbPb}{\ensuremath{\mbox{Pb--Pb}}}
\newcommand{\GeV}{\ensuremath{\mathrm{GeV}\kern-0.05em}}
\newcommand{\GeVc}{\ensuremath{\mathrm{GeV}\kern-0.05em/\kern-0.02em c}}
\newcommand{\sqrts}{\ensuremath{\sqrt{s_{\mathrm{NN}}}}}
\newcommand{\pT}{\ensuremath{p_{\mathrm{T}}}}
\newcommand{\RL}{\ensuremath{R_{\mathrm{L}}}}
\newcommand{\pTi}{\ensuremath{p_{\mathrm{T},i}}}
\newcommand{\pTtrack}{\ensuremath{p_{\mathrm{T,track}}}}
\newcommand{\sigmaeec}{\ensuremath{\Sigma_{\mathrm{EEC}}}}

\newcommand{\kT}{\ensuremath{k_{\mathrm{T}}}}
\newcommand{\pThard}{\ensuremath{p_{\mathrm{T,hard}}}}
\newcommand{\etajet}{\ensuremath{\eta_{\mathrm{jet}}}}
\newcommand{\pTjet}{\ensuremath{p_{\mathrm{T}}^{\mathrm{jet}}}}
\newcommand{\pTchjet}{\ensuremath{p_{\mathrm{T}}^{\mathrm{ch\; jet}}}}
\newcommand{\pTfulljet}{\ensuremath{p_{\mathrm{T}}^{\mathrm{full\; jet}}}}
\newcommand{\pTtruth}{\ensuremath{p_{\mathrm{T,truth}}^{\mathrm{ch\; jet}}}}
\newcommand{\pTdet}{\ensuremath{p_{\mathrm{T,det}}^{\mathrm{ch\; jet}}}}
\newcommand{\Nevent}{\ensuremath{N_{\mathrm{event}}}}
\newcommand{\Ninc}{\ensuremath{N_{\mathrm{jets}}}}
\newcommand{\Sinc}{\ensuremath{\sigma_{\mathrm{jets}}}}

\title{Medium modification of the charge-weighted Energy-Energy Correlators in Pb+Pb collisions at $\sqrt{s_{NN}}=5.02$~TeV}

\date{\today  \hspace{1ex}}

\author{Han-Yao Liu\footnote{These authors contributed equally to this work}}
\affiliation{School of Mathematics and Physics, China University of Geosciences, Wuhan 430074, China}

\author{Shi-Yong Chen\footnote{These authors contributed equally to this work}}
\affiliation{Huanggang Normal University, Huanggang 438000, China}
\affiliation{Key Laboratory of Quark \& Lepton Physics (MOE) and Institute of Particle Physics, Central China Normal University, Wuhan 430079, China}


\author{Wei Dai\footnote{weidai@mail.cug.edu.cn}}
\affiliation{School of Mathematics and Physics, China University of Geosciences, Wuhan 430074, China}

\author{Ben-Wei Zhang}
\affiliation{Key Laboratory of Quark \& Lepton Physics (MOE) and Institute of Particle Physics, Central China Normal University, Wuhan 430079, China}

\begin{abstract}

We report a systematic study of medium-induced modifications in charge-dependent jet substructure using the charge-weighted Energy-Energy Correlators (EEC) in p+p and $0-10\%$ central Pb+Pb collisions at $\sqrt{s_{\mathrm{NN}}} = 5.02\ \mathrm{TeV}$. Charged hadron jets as well as flavor-separated quark and gluon-initiated jets with momentum 40–60 GeV and $R=0.4$ are analyzed. The ratio of the charge-weighted distribution to the inclusive EEC, which reflects the magnitude of charge correlations, is consistently negative, demonstrating the dominance of opposite-charge pairs due to charge conservation.  In p+p collisions, Pythia simulations predict a maximum opposite-charge correlations at small relative angular distance $R_\text{L}$  of charge hadron pairs that remains constant; however, they fail to describe the experimentally observed decorrelations as $R_\text{L}$ decreases.  A clear flavor dependence is presented: gluon-initiated jets exhibit less opposite-charge correlations in the transition and smaller $R_\text{L}$ region than quark-initiated jets, while more opposite-charge correlations in the larger $R_\text{L}$ region. In Pb+Pb collisions, significant medium modifications are observed. The A+A to p+p ratio for charge correlations shows a universal pattern across jet types, shows no flavor dependence: 
jet quenching enhances opposite-charge correlations at small angles while reducing it at large angles, leading to a steeper $R_\text{L}$ dependence of charge correlations in A+A, indicating a more rapid decorrelated behavior when $R_\text{L}$ increases. A distinctive V-shaped modification pattern appears in the plateau-like enhancement in the transition and small-$R_\text{L}$ region, independent of jet flavor. Through factorization of the EEC into charged hadron pair multiplicity and average energy weighting distributions, we identify enhanced energy weighting of opposite-charge pairs at small $R_\text{L}$ as the origin of the V-shaped modification. Also, the plateau-like enhancement of charge correlations is found to be unrelated to selection bias effects.

\end{abstract}


\maketitle

\section{Introduction}

In high-energy heavy-ion collisions, a new state of matter—the quark-gluon plasma (QGP)—can be created, whose properties can be indirectly probed and studied through modifications experienced by hard probes traversing it~\cite{Gyulassy:1990ye,Luo:2017faz,Tang:2020ame,Zhang:2021xib,Shou:2024uga}. This phenomenon is named as jet quenching. A prominent hard probe is the high-energy partons, or jets, produced in an initial hard scattering. Its evolution generally proceeds as follows: first, the high-transverse-momentum parton generated in the hard collision carries a certain virtuality and undergoes initial splittings in the vacuum, described by perturbative QCD via DGLAP evolution~\cite{Hoche:2017hno}. Subsequently, the process becomes increasingly non-perturbative, with further splittings occurring at progressively smaller angles—commonly defined as parton showering—until the partons hadronize and are reconstructed as jets via clustering algorithms.

Studying the modifications of jets as they traverse through the QGP has long been a central topic in the field. In recent years, sustained investigation of jet substructure has deepened our understanding of its modification and evolution in the hot and dense medium. Jet substructure observables, constructed from final-state hadrons, encode information about parton shower evolution and hadronization~\cite{ATLAS:2018bvp}. A class of such observables, specifically designed to constrain nonperturbative processes, has been used to study modifications to the perturbative phase of jet evolution in the medium. Another category of substructure observables enables the separation of perturbative and nonperturbative regimes across the full angular range—particularly shedding light on the hadronization process~\cite{Larkoski:2017jix,Liu:2022kzv,Zhang:2023jpe,Kogler:2018hem,Dai:2012am,Dai:2018mhw,Luo:2018pto,Wang:2019xey}. The two-point energy-energy correlators(EEC)~\cite{Basham:1977iq,Basham:1978bw,Basham:1978zq,PLUTO:1985yzc} emerges as one of them, which characterizes the energy flow within jets via 1→2 parton splittings, exhibits a distribution separable into three distinct angular regimes: (i) small angles dominated by the free diffusion of hadrons, (ii) large angles governed by perturbative parton shower evolution, and (iii) an intermediate transition region between them~\cite{Chen:2020vvp,STAR:2025jut,ALICE:2024dfl,ALICE:2025igw,Mazzilli:2024ots,Liang-Gilman:2025gjl,CMS:2025ydi,CMS:2024mlf}. In recent years, the EEC has been explored from multiple aspects as a probe of jet quenching phenomena~\cite{Devereaux:2023vjz,Fu:2024pic,Barata:2024wsu,Nambrath:2025ttz,Shen:2024oif,Barata:2023vnl,Chen:2024quk,Barata:2024ukm,Bossi:2025kac,Barata:2025jhd,Xing:2024yrb,Andres:2023ymw,Barata:2025uxp,Andres:2022ovj,Andres:2023xwr,Andres:2024ksi,Yang:2023dwc,Bossi:2024qho,Andres:2024xvk,Apolinario:2025vtx,Andres:2024hdd,Andres:2024pyz,Barata:2025fzd}.
In heavy-ion collisions, it further provides insights into initial-state cold nuclear matter effects~\cite{Devereaux:2023vjz,Fu:2024pic,Barata:2024wsu,Nambrath:2025ttz,Shen:2024oif}, parton energy loss mechanisms~\cite{Barata:2023vnl,Chen:2024quk}, medium response~\cite{Barata:2024ukm,Bossi:2025kac,Barata:2025jhd}, the properties of in-medium light-flavor quarks~\cite{Xing:2024yrb,Chen:2024quk}, mass effects of heavy-flavor quarks~\cite{Andres:2023ymw,Shen:2024oif,Barata:2025uxp,Mazzilli:2024ots}, and the physical properties of the quark-gluon plasma (QGP)~\cite{Andres:2022ovj,Andres:2023xwr,Andres:2024ksi,Yang:2023dwc,Bossi:2024qho,Andres:2024xvk,Apolinario:2025vtx,Andres:2024hdd,Andres:2024pyz,Barata:2025fzd,Liang-Gilman:2025gjl}.

The EEC can be categorized by considering hadron pairs carrying like- or opposite-charges, enabling the study of charge correlations during hadronization and providing constraints on hadronization models~\cite{Lee:2023npz,Chien:2021yol}. This is achieved by including the charge product in the energy weighting, yielding the charge-weighted EEC. Furthermore, the ratio of the charge-weighted EEC to the inclusive EEC offers a reweighted measure of charge correlations as a function of hadron-pair angular separation. They have garnered significant attention from experimental collaborations such as STAR and ALICE~\cite{STAR:2025jut,C3NT2024}. Owing to its sensitivity to timescale separation within jets, a detailed investigation of the modification patterns of these observables in A+A collisions becomes essential. Such studies not only help probe and track the jet evolution process in the hot and dense medium~\cite{Mazzilli:2024ots,Marchesini:1994wr,Komiske:2022enw,Neill:2022lqx}, but also establish a valuable baseline for future exploration of hadronization mechanisms in heavy-ion collisions. 

In this work, we present in Chapter II the theoretical baseline obtained with Pythia and compare it with experimental results from the RHIC and the LHC. In Chapter III, we provide for the first time theoretical predictions for the charge-weighted EEC distribution of charged hadron jets with $40-60$~GeV and $R=0.4$ in $0-10\%$ central Pb+Pb collisions, and discuss the jet quenching effects observed. In Chapter IV, we compute the theoretical predictions for the $R_\text{L}$ distribution of the ratio of the charge-weighted EEC to the inclusive EEC and analyze the influence of hot and dense medium modifications. Finally, in Chapter V, we summarize our conclusions and outline future perspectives.

\section{Charged weighted EEC/inclusive EEC ratios in p+p Collisions at the RHIC and the LHC}

The charge-weighted EEC for final-state charged hadrons within jets is constructed EEC the same way as the previous~\cite{ALICE:2024dfl,Chen:2024quk}, but including the charge product of the hadron pairs in the energy weighting~\cite{STAR:2025jut}. It is experimentally expressed as follows:
\begin{widetext}
\begin{eqnarray}
Charge-weighted\ EEC
&=& \frac{1}{ N_{\text{jet}}\cdot\Delta R}\int_{R_{\text{L}}-\frac{1}{2}\Delta R}^{R_{\text{L}}+\frac{1}{2}\Delta R}\sum_{\text{jets}}\sum_{i, j}\frac{Q_iQ_jp_{\text{T}, i}p_{\text{T}, j}}{p^2_{\rm T, jet}}\delta(R_{\text{L}}'-R_{\text{L}, ij})\ R_{\text{L}} \\
&=& EEC_\text{like} - EEC_\text{opposite}
\label{EEC-equ}
\end{eqnarray}
\end{widetext}
where $i$ and $j$ denote two particles within a jet, $R_{L,ij} = \sqrt{(\phi_j - \phi_i)^2 + (\eta_j - \eta_i)^2}$ is their angular separation, $Q_i$ and $Q_j$ are their electric charges, $p_{\text{T}, i}/p_{T,\text{jet}}$ and $p_{\text{T}, j}/p_{T,\text{jet}}$ are their transverse momentum fractions relative to the jet. $\Delta R$ is the angular bin width and $N_{\text{jet}}$ is the total number of jets. It can naturally separate the EEC into two components: one constructed using only like-charge pairs and one using only opposite-charge pairs. When considering the sign of the charge pairs $Q_iQ_j$, the charge-weighted EEC corresponds to the contribution from like-charge pairs minus the contribution from opposite-charge pairs.

\begin{figure}[!htb]
\centering

\includegraphics[scale=0.32]{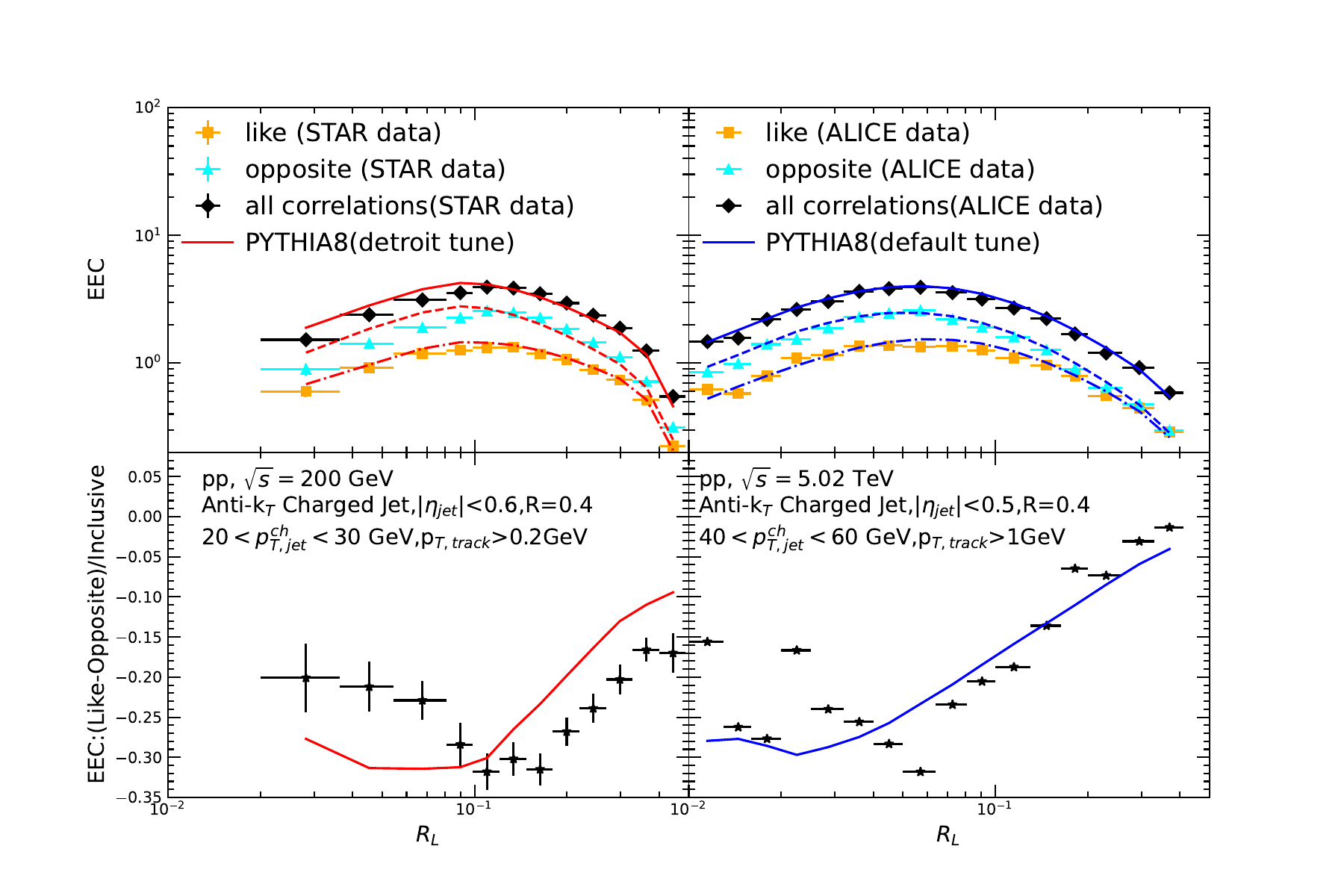}
\vspace{-1cm}
\caption{PYTHIA8 simulation results of normalized charge-weighted EEC with its like- (opposite-) charge components (upper panel) and charge-weighted EEC/inclusive EEC ratio distributions (bottom panel) as functions of $R_{\rm L}$ for inclusive charged jets with a jet size of $R=0.4$ produced in $\rm p+p$ collisions at $\sqrt{s}=200$~GeV(left panel) and $\sqrt{s}=5.02$~TeV(right panel). The results  are compared with the STAR~\cite{STAR:2025jut} and ALICE experimental data~\cite{C3NT2024}, respectively.}
\label{fig:baseline}
\end{figure}

We used the Monte Carlo event generator \textsc{Pythia} v8.313~\cite{Sjostrand:2014zea} to simulate p+p collisions at LHC energy $\sqrt{s}=5.02$~TeV with the default tune~\cite{Skands:2014pea} and RHIC energy $\sqrt{s}=200$~GeV with the Detroit tune~\cite{Aguilar:2021sfa}, which is the same as the STAR collaboration implemented in~\cite{STAR:2025jut}. Jets were reconstructed using the anti-$k_T$ algorithm from the \textsc{FastJet} v3.4.0 package~\cite{Cacciari:2008gp}, with a jet radius parameter $R = 0.4$. The selected jets have transverse momenta $20\ \text{GeV/c} <p_{\text{T,jet}} < 30\ \text{GeV/c}$, jet rapidity $|\eta_{\text{jet}}| < 0.6$, are composed of charged particles with $p_{T,had} >0.2\ \text{GeV/c}$ at $\sqrt{s}=200$~GeV, and have transverse momenta $40\ \text{GeV/c} <p_{\text{T,jet}} < 60\ \text{GeV/c}$, jet rapidity $|\eta_{\text{jet}}| < 0.5$, are composed of charged particles with $p_{T,had} >1\ \text{GeV/c}$ at $\sqrt{s}=5.02$~TeV.

We computed, in p+p collisions at both $\sqrt{s}=200$~GeV and $5.02$~TeV, the distributions of the charge-weighted EEC in the upper panel of Fig.~\ref{fig:baseline}, as well as the ratio of the charge-weighted EEC to the inclusive EEC in the bottom panel. The like-charge component (dashed line), the opposite-charge component (dash-dotted line), and the inclusive EEC distributions (solid line) are plotted together for each collision energy, respectively. They are confronted with the ALICE~\cite{C3NT2024} and STAR experimental data~\cite{STAR:2025jut}, respectively. We find that for both cases, the like-charge components agree well with the experimental data across the entire $R_\text{L}$ range. The results of the opposite-charge component underestimate the experimental data at larger $R_\text{L}$ and overestimate that at smaller $R_\text{L}$. Its peak is shifted towards smaller angles, which leads to the final results of the inclusive distributions shifting to larger angles and displaying over-prediction of the experimental data at smaller $R_\text{L}$. 

In the lower panel of Fig.~\ref{fig:baseline}, we present the ratio of the charge-weighted distribution to the inclusive EEC at two collision energies, $200$~GeV and $5.02$~TeV. This ratio quantifies the magnitude of charge correlations across the full range from $-1$ to $+1$. For the jets in our studied dynamic range, the distribution is predominantly concentrated at negative values. A ratio of zero corresponds to the extreme scenario of an infinite thermal bath, where the probabilities of forming like-charge and opposite-charge pairs are equal. However, within our studied range (with an averaged jet charge $Q=0.006$ and limit particle numbers within a jet), the opposite-charge correlations are observed to be stronger than the like-charge correlations, both overall and prevalently—a consequence of charge conservation. This observable precisely reflects the relative abundance distribution of these two types of correlations in terms of relative angular distance ($R_\text{L}$), with a more negative value corresponding to increased correlations of opposite-charge pairs.

In the larger $R_\text{L}$ region beyond the transition, the positive slope of the ratio with increasing $R_\text{L}$ indicates that opposite-charge correlations become decorrelated in the shower-dominated regime. This feature is also captured by Pythia simulations, whose predicted slope agrees with the experimental data. Conversely, in the small $R_\text{L}$ region below the transition—dominated by free diffusion of hadrons—the opposite-charge correlations are strongest and remain relatively flat as a function of $R_\text{L}$. Intriguingly, we are unable to account for the experimentally observed negative slope in this region using Pythia, which suggests the emergence of opposite-charge decorrelations as the $R_\text{L}$ distance decreases.

\section{Charged weighted EEC in Pb+Pb Collisions}

In this work, we examine the impact of jet quenching on the charge correlations—as revealed by the ratio of the charge-weighted EEC to the inclusive EEC—in a two-step approach. We first compute the A+A to p+p ratio of the charge-weighted EEC within the same transverse momentum interval as the p+p baseline.

We employ the \textsc{SHELL} model, a Monte Carlo framework, to simulate the energy loss processes of partons produced from initial hard scatterings as they traverse the Quark-Gluon Plasma (QGP) in heavy-ion collisions. The \textsc{SHELL} model is primarily based on a Langevin transport equation and incorporates three key physical processes: elastic collisional energy loss, in-medium gluon radiation, and the medium response. This model has demonstrated a good description of a wide range of experimental observables, including those related to jet quenching~\cite{Yan:2020zrz,Chen:2022kic,Dai:2022sjk,Li:2022tcr,Wang:2024plm,Li:2024pfi}.

The model initializes the dynamical distribution of partons using a Glauber-model-based nuclear geometry~\cite{Alver:2008aq}. These partons are then propagated step-by-step through a QGP medium whose evolution is described by relativistic hydrodynamics. At each evolution time step $\Delta t$, a parton may simultaneously undergo elastic scattering, radiate gluons, and induce a medium response. The elastic energy loss per unit length is calculated using the Hard Thermal Loop (HTL) formula\cite{Neufeld:2010xi}:
\begin{equation}
    \frac{dE}{dt} = \frac{\alpha_s C_s}{2} \mu_D^2 \ln\left(\frac{\sqrt{ET}}{\mu_D}\right),
\end{equation}
where $\alpha_s$ is the running coupling constant of the strong interaction, taken as 0.3 in our calculations, $C_s$ is the Casimir operator ($C=3$ for a gluon, $C=4/3$ for a quark), $\mu_D$ is the Debye screening mass, and $E$ and $T$ are the parton energy and local medium temperature, respectively.

The energy deposited into the medium via these elastic collisions accumulates and interacts with the QGP, eventually hadronizing upon freeze-out to form ``wake'' particles. This constitutes the medium response mechanism in \textsc{SHELL}. The momentum distribution of these wake hadrons is calculated using the Cooper-Frye prescription with perturbations~\cite{Cooper:1974mv,Casalderrey-Solana:2016jvj}:
\begin{align}
  \label{eq:wake}
  E\frac{\dd \Delta N}{\dd^3p}=&\frac{1}{32 \pi} \, \frac{m_{\rm T}}{T^5} \, \cosh(y-y_j) \exp\left[-\frac{m_{\rm T}}{T}\cosh(y-y_j)\right] \notag \\
      &\times \Big\{ p_T \Delta P_{\rm T} \cos (\phi-\phi_j) \notag \\
      &+\frac{1}{3}m_{\rm T} \, \Delta M_{\rm T} \, \cosh (y-y_j) \Big\},
\end{align}
Here, $E$, $\mathbf{p}$, $\Delta N$, $m_T$, $p_T$, $y$, and $\phi$ are the energy, momentum, yield, transverse mass, transverse momentum, rapidity, and azimuthal angle of the wake hadron, respectively. $y_j$ and $\phi_j$ are the rapidity and azimuthal angle of the energy-depositing parton. $\Delta P_T$ and $\Delta M_T = \Delta E / \cosh(y_j)$ represent the transverse momentum and transverse mass transferred to the medium, with $\Delta E$ being the accumulated deposited energy. The first and second terms on the right-hand side correspond to the calculations of the momentum and energy of the wake hadrons, respectively.

Within \textsc{SHELL}, the probability for a parton to radiate a gluon in a time step $\Delta t$ within the QGP medium is given by:
\begin{equation}
   P_{rad}(t,\Delta t)=1-e^{-\left\langle N(t,\Delta t)\right\rangle} \, .
\end{equation}
where $\langle N(t, \Delta t)\rangle$ is the average number of radiated gluons, calculated from the medium-induced gluon spectrum within the Higher-Twist (HT) method\cite{Majumder:2009ge,Guo:2000nz,Zhang:2003yn,Cao:2017hhk}:
\begin{equation}
    \frac{\dd N}{\dd x \dd k^{2}_{\perp}\dd t}=\frac{2\alpha_{s}C_sP(x)\hat{q}}{\pi k^{4}_{\perp}}\sin^2(\frac{t-t_i}{2\tau_f})(\frac{k^2_{\perp}}{k^2_{\perp}+x^2M^2})^4
\end{equation}
Here, $k_\perp$ is the transverse momentum of the radiated gluon, $x = E_g/E_p$ is the energy fraction of the gluon relative to the parent parton, M is the mass of the parent parton, $P(x)$ is the vacuum QCD splitting function, $\tau_f = 2Ex(1-x)/(k_\perp^2 + x^2 M^2)$ is the gluon formation time, $t - t_i$ is the time interval in the parton frame, and $\hat{q} = \hat{q}_0 (T/T_0)^3 (p\cdot u)/E$ is the jet transport parameter, where $\hat{q}_0 = 1.5~\text{GeV}^2/\text{fm}$, $u^\mu$ is the four-velocity of the QGP at the given space-time point, and $T_0$ is the initial temperature. This transport parameter is the key dynamical quantity governing the strength of parton energy loss. The number of radiated gluons in a given step is sampled from a Poisson distribution, and their four-momenta are subsequently calculated.

The space-time evolution of the expanding QGP fireball is provided by the (3+1)-dimensional CLVisc hydrodynamic model\cite{Shen:2014vra}. The hadronization temperature is set to $T_c = 165$ MeV; partons and the deposited wake energy that fall below this temperature undergo hadronization. The final-state hadrons are obtained by first forming color-singlet strings using the colorless method from the JETSCAPE collaboration~\cite{Putschke:2019yrg}, followed by hadronization and hadron decays via the PYTHIA Lund string model.

\begin{figure}[!htb]
\centering
\includegraphics[scale=0.6]{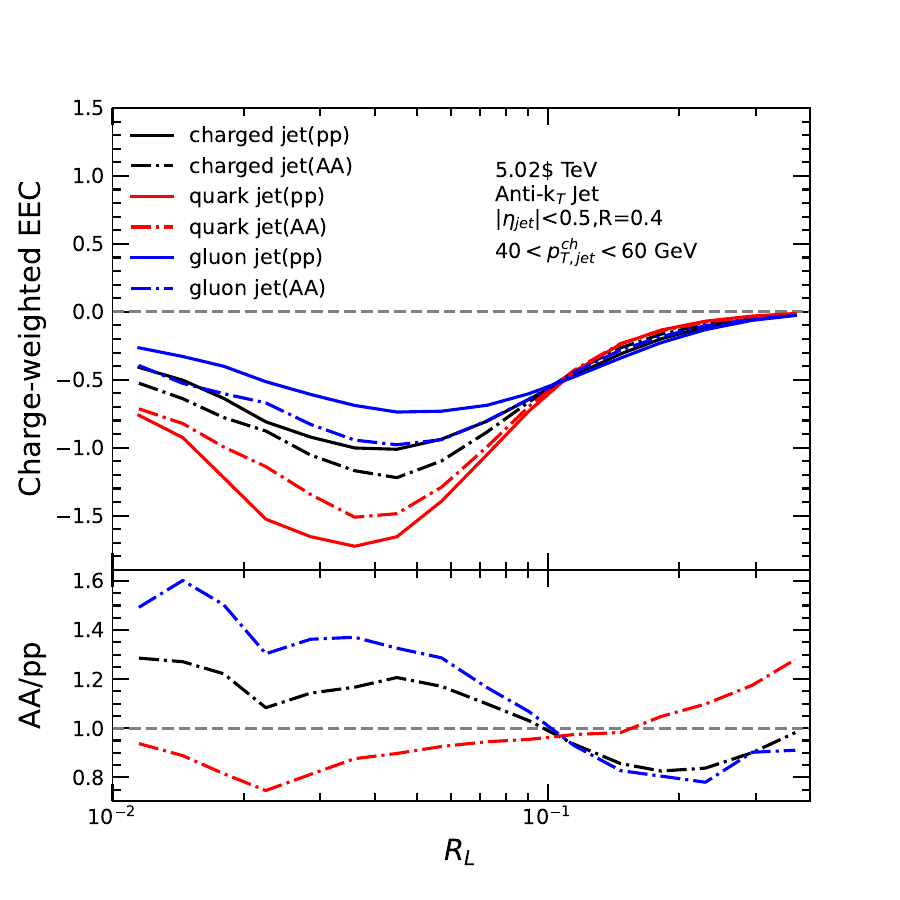}
\vspace{-1.0 cm}
\caption{Charge-weighted EEC in p+p and central $(0-10\%)$ Pb+Pb collisions and their A+A/p+p ratios distributions as functions of $R_\text{L}$ for inclusive charged, gluon and quark jets with a radius of $R=0.4$ in jet $p_{\rm T}$ interval $40-60$~GeV at $\sqrt{s}=5.02$~TeV.} 
\label{fig:ceecAA}
\end{figure}

Based on Eq.~\ref{EEC-equ}, we note that the charge-weighted EEC is numerically equivalent to the like-charge component minus the opposite-charge component of the inclusive EEC. Our previous work~\cite{Chen:2024quk} further demonstrated that the in-medium modification of the EEC exhibits characteristically distinct behaviors between gluon- and quark-initiated jets. We demonstrate in Fig.~\ref{fig:ceecAA} computed the charge-weighted distributions of EEC in p+p and A+A (upper panel) and their $\mathrm{A+A/p+p}$ ratios (bottom panel) for inclusive charged jets, quark-initiated, and gluon-initiated jets in central (0--10\%) Pb+Pb collisions at $\sqrt{s_{\mathrm{NN}}} = 5.02$~TeV. All jets reconstructed from charged particles with $p_{T,had} >1\ \text{GeV/c}$ with the anti-$k_T$ algorithm using a radius parameter of $R = 0.4$, and are selected with transverse momentum in the range of 40--60 GeV/$c$ and pseudo-rapidity $|\eta^{\text{jet}}| < 0.5$.

In the p+p system, the charge-weighted EEC distribution lies predominantly in the negative region under the studied conditions, indicating that opposite-charge pairs contribute more than like-charge pairs. Beyond the transition region toward larger $R_\text{L}$, the distribution exhibits a positive slope. In contrast, toward smaller $R_\text{L}$ it shows a negative slope—that is, the difference between opposite- and like-charge contributions decreases as $R_\text{L}$ decreases. Regarding flavor dependence, the charge-weighted distribution of gluon-initiated jets lies closer to zero than that of quark-initiated jets. In the small-$R_\text{L}$ region, the difference between opposite- and like-charge pairs is significantly smaller in gluon jets than in quark jets, while in the large-$R_\text{L}$ region, this difference becomes slightly larger in gluon jets than in quark jets.  In A+A, we observe that their behavior of $\mathrm{A+A/p+p}$ ratios resembles the previously reported modification of the EEC in a hot dense medium, with characteristic modification patterns for quark and gluon jets also being consistent~\cite{Chen:2024quk}. This suggests a common modification mechanism for the EEC distributions: the enhancement in the small-$R_\text{L}$ region is dominated by selection bias and the initial spectrum properties of gluon jets, while the suppression at larger $R_\text{L}$ that increases with $R_\text{L}$ is primarily driven by broadening effects due to radiative energy loss. Then, in the small-$R_\text{L}$ region around $R \approx 0.02$, we observe a characteristic V-shaped modification pattern. To further investigate the origin of this feature, it is necessary to examine the medium modifications of the like-charge and opposite-charge components separately.

\begin{figure}[!htb]
\centering
\includegraphics[scale=0.6]{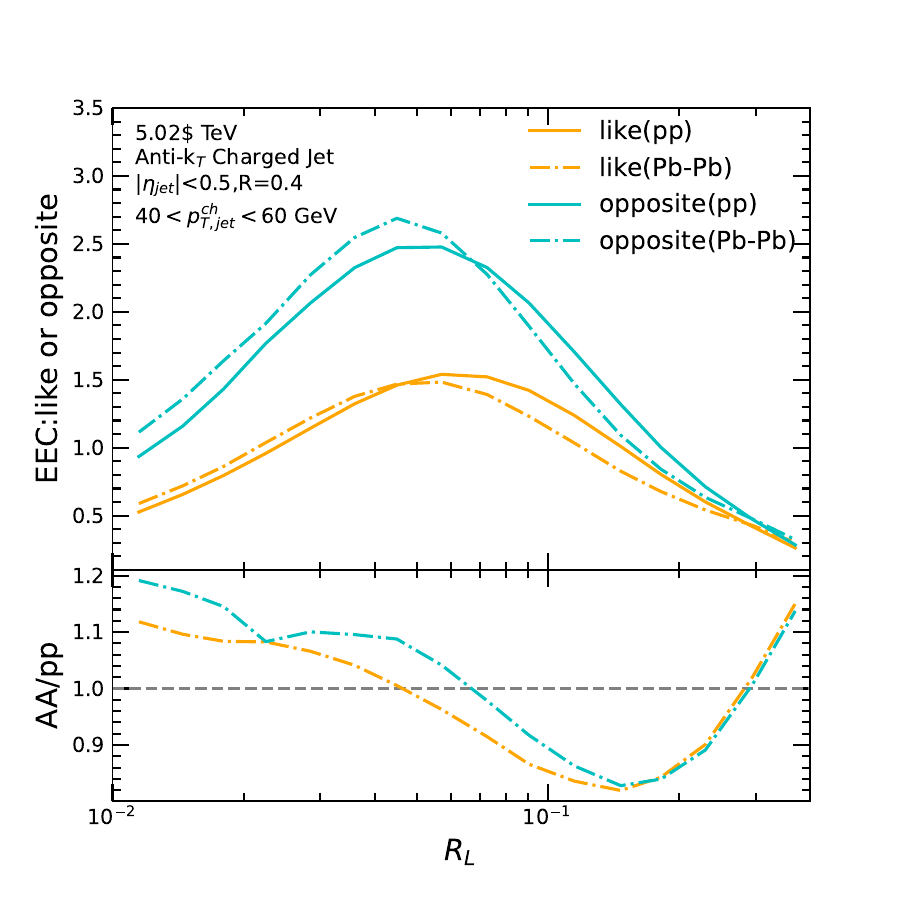}
\vspace{-1.cm}
\caption{ Like and opposite components of EEC in p+p and central $(0-10\%)$ Pb+Pb collisions and their A+A/p+p ratios distributions as functions of $R_\text{L}$ for inclusive charged jets with a radius of $R=0.4$ at $\sqrt{\rm s}=5.02$~TeV in jet $p_{\rm T}$ interval $40-60$~GeV.}
\label{fig:compo}
\end{figure}

In Fig.~\ref{fig:compo}, we present the distributions of both like-charge and opposite-charge components in p+p and A+A collisions (upper panel), with their corresponding A+A to p+p ratios shown in the bottom panel. While their modification patterns are generally similar to those of the inclusive EEC, the opposite-charge component exhibits a stronger enhancement in the transition and small $R_\text{L}$ region compared to the like-charge part. Moreover, the V-shaped modification structure at the small $R_\text{L}$ is found exclusively in the opposite-charge component. This observation is consistent with the enhanced A+A to p+p ratio and the V-shaped modification pattern observed in the charge-weighted EEC within the small-$R_\text{L}$ region.

To further investigate the raw charge correlations without energy weighting and to understand how jet quenching manifests, we adopt the same methodology as in previous work~\cite{Chen:2024quk}: we factorize the EEC distribution into the distribution of the average number of final-state charged hadron pairs per jet as a function of  $R_\text{L}$, and the corresponding average energy-energy weighting of hadron pairs within each  $R_\text{L}$ bin $\langle \text{weight} \rangle$. The hadron pairs can then be categorized according to their like-charge or opposite-charge nature. We therefore decompose the EEC distribution for the like-charge and the opposite-charge components as follows:
\begin{eqnarray}
\Sigma^{\text{comp}}_{\text{EEC}} = \frac{N^{\rm total}_{\rm pair}}{N_{\rm jet}} \thinspace \cdot
\frac{\Delta N^{\text{comp}}_{\rm pair}}{N^{\rm total}_{\rm pair} \Delta R} \thinspace \cdot \langle \rm weight \rangle^{\text{comp}} 
\end{eqnarray}
where $N^{\text{total}}_{\text{pair}}$ is the total number of particle pairs, $\Delta N_{\text{pair}}$ is the number of pairs within an angular bin of width $\Delta R$ centered at $R$, and $\langle \text{weight} \rangle$ represents the average energy-weight carried by pairs in that bin. 
\begin{figure}[!htb]
\centering
\includegraphics[scale=0.6]{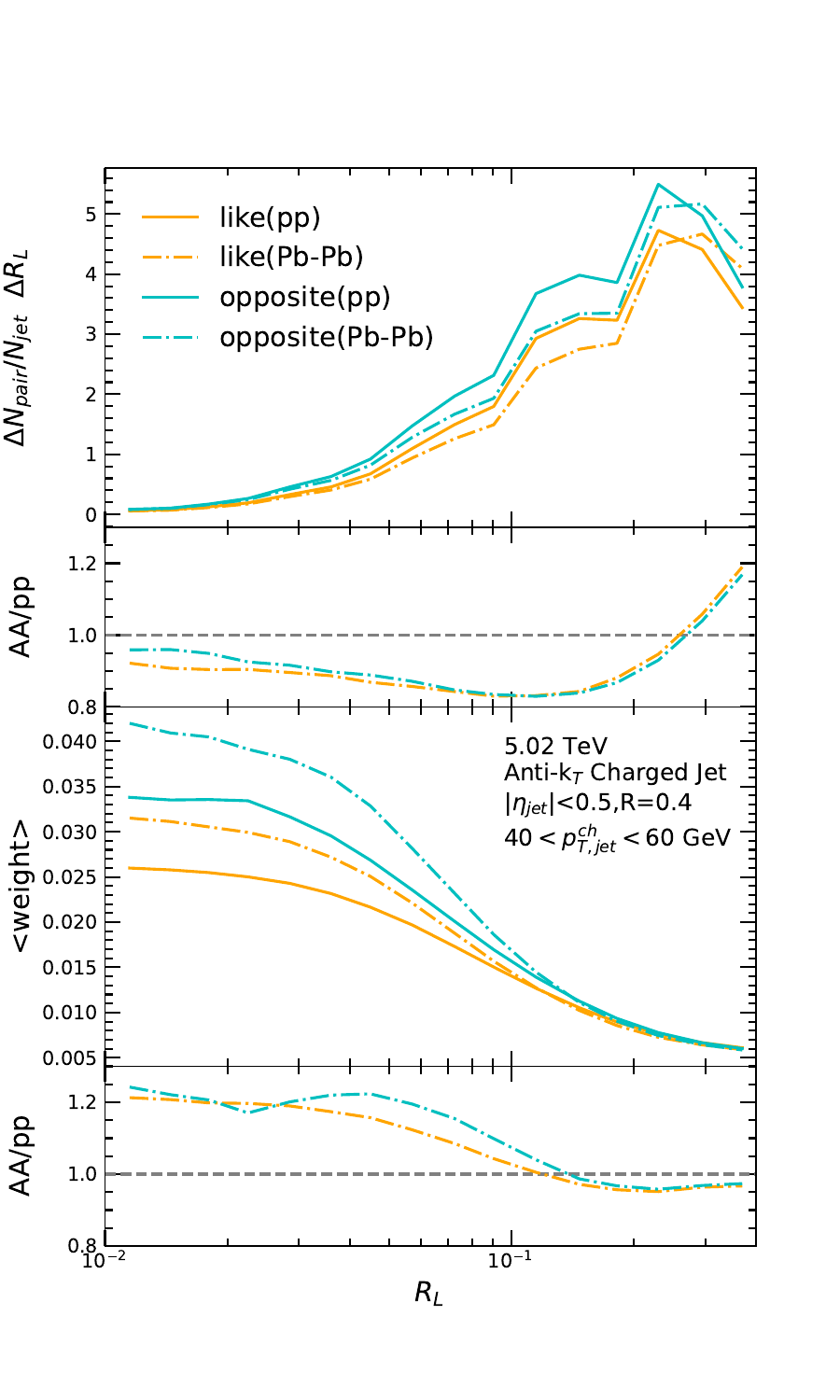}
\vspace{-1.5cm}
\caption{ Number of jets normalized pair-number distributions (upper panel) and averaged energy weight (lower panel) distributions of like and opposite components EEC in p+p and central $(0-10\%)$ Pb+Pb collisions and their A+A/p+p ratios distributions as functions of $R_\text{L}$ for inclusive charged jets with a radius of $R=0.4$ at $\sqrt{\rm s}=5.02$~TeV in jet $p_{\rm T}$ interval $40-60$~GeV.}
\label{fig:factoriz}
\end{figure}

In the upper panel of Fig.~\ref{fig:factoriz}, the $R_\text{L}$ distributions of like-charge and opposite-charge hadron pairs in p+p collisions are shown as solid lines. Without energy weighting, the raw hadron pairs are predominantly distributed at large $R_\text{L}$ values. As ensured by charge conservation, the number of opposite-charge pairs exceeds that of like-charge pairs across the entire $R_\text{L}$ range. In A+A collisions, the $R_\text{L}$ distributions of both types of hadron pairs shift toward larger $R_\text{L}$, though the suppression of opposite-charge pairs at small angles is slightly weaker than that of like-charge pairs, and the enhancement at large angles is also somewhat milder for opposite-charge pairs. In the lower panel of Fig.~\ref{fig:factoriz}, we observe that the average energy weighting of opposite-charge pairs is significantly larger than that of like-charge pairs in both the transition region and at small $R_\text{L}$. While this difference becomes modest at large $R_\text{L}$, it is accentuated by the contrasting angular distributions of hadron pairs in this region. In A+A case, the opposite-charge pairs exhibit a stronger enhancement relative to p+p collisions at small $R_\text{L}$ compared to like-charge pairs, whereas the suppression at large $R_\text{L}$ is nearly identical for both types. Notably, a characteristic V-shaped modification pattern emerges in the opposite-charge component, manifested in the pronounced enhancement at the small-$R_\text{L}$ region. It can be identified as the underlying origin of the V-shaped modification pattern observed in the charge-weighted EEC.

\section{charged weighted EEC/inclusive EEC ratios in Pb+Pb Collisions}

Finally, we turn to the interpretation of the ratio of the charge-weighted distribution to the inclusive EEC. As noted earlier, this ratio quantitatively characterizes the degree of charge correlations, allowing us to focus specifically on charge correlations among hadron pairs. Since the $R_\text{L}$ variable of the EEC observable can be used to infer the jet formation process inversely, Fig.~\ref{fig:baseline} illustrates that as the jet evolves from the large-$R_\text{L}$ region—dominated by QCD shower processes—toward the transition region, the opposite-charge correlations strengthens relative to like-charge correlations. Proceeding further into the hadronization stage (from the transition toward small $R_\text{L}$), where hadron pairs are formed at minimal angular separation, the opposite-charge correlations becomes the strongest. We now proceed to evaluate the impact of jet quenching on this charge correlations.
\begin{figure}[!htb]
\centering
\includegraphics[scale=0.6]{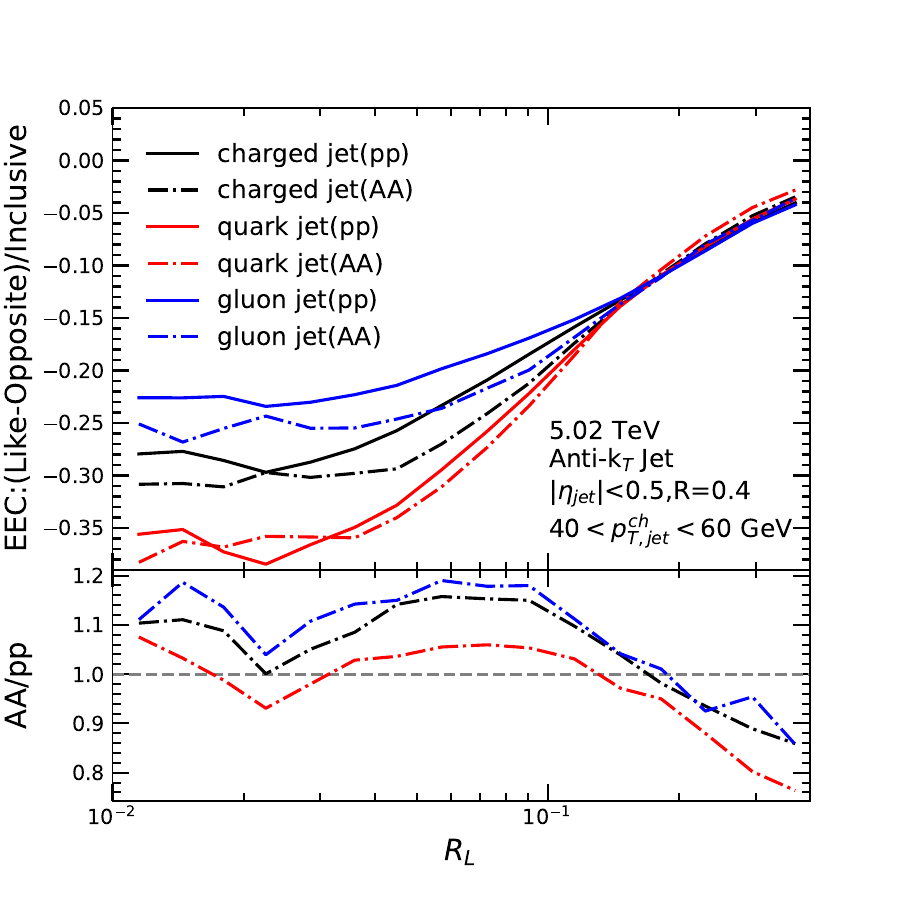}
\vspace{-1cm}
\caption{ Charge weighted EEC/inclusive EEC ratios in p+p and central $(0-10\%)$ Pb+Pb collisions and their A+A/p+p ratios distributions as functions of $R_{L}$ for inclusive charged, gluon and quark jets with a radius of $R=0.4$ in jet $p_{\rm T}$ interval $40-60$~GeV at $\sqrt{s}=5.02$~TeV.}
\label{fig:aappall}
\end{figure}

In. Fig.~\ref{fig:aappall}, we present the charge correlations and their A+A/p+p ratios for charged hadron jets, gluon-initiated jets, and quark-initiated jets with a momentum of $40–60$~GeV and $R=0.4$ in $0–10\%$ Pb+Pb collisions at $5.02$~TeV. In the p+p baseline, the slope of the charge correlations as a function of $R_\text{L}$ is steeper for quark jets than for gluon jets. Consequently, quark jets exhibit a quicker opposite-charge de-correlations at large $R_\text{L}$, while displaying the largest opposite-charge correlations at small $R_\text{L}$. 
The A+A/p+p ratios for all three jet types show a consistent modification pattern: a plateau-like enhancement in the transition and small-$R_\text{L}$ region, and a suppression that increases with $R_\text{L}$ in the large-$R_\text{L}$ region. This demonstrates that jet quenching reduces opposite-charge correlations at large $R_\text{L}$ while enhancing it in the transition and small-$R_\text{L}$ region, resulting in an increased slope of the charge correlations with $R_\text{L}$. Two important observations emerge: 1. The characteristic modification patterns that distinguish quark jets from gluon jets in the charge-weighted EEC and inclusive EEC are canceled out when taking their ratio; 2. The distinctive V-shaped medium modification observed in the A+A/p+p ratio of the opposite-charge component of the charge-weighted EEC appears at the same $R_\text{L}$ position, independent of the jet initiator flavor. 

\begin{figure}[!htb]
\centering
\includegraphics[scale=0.6]{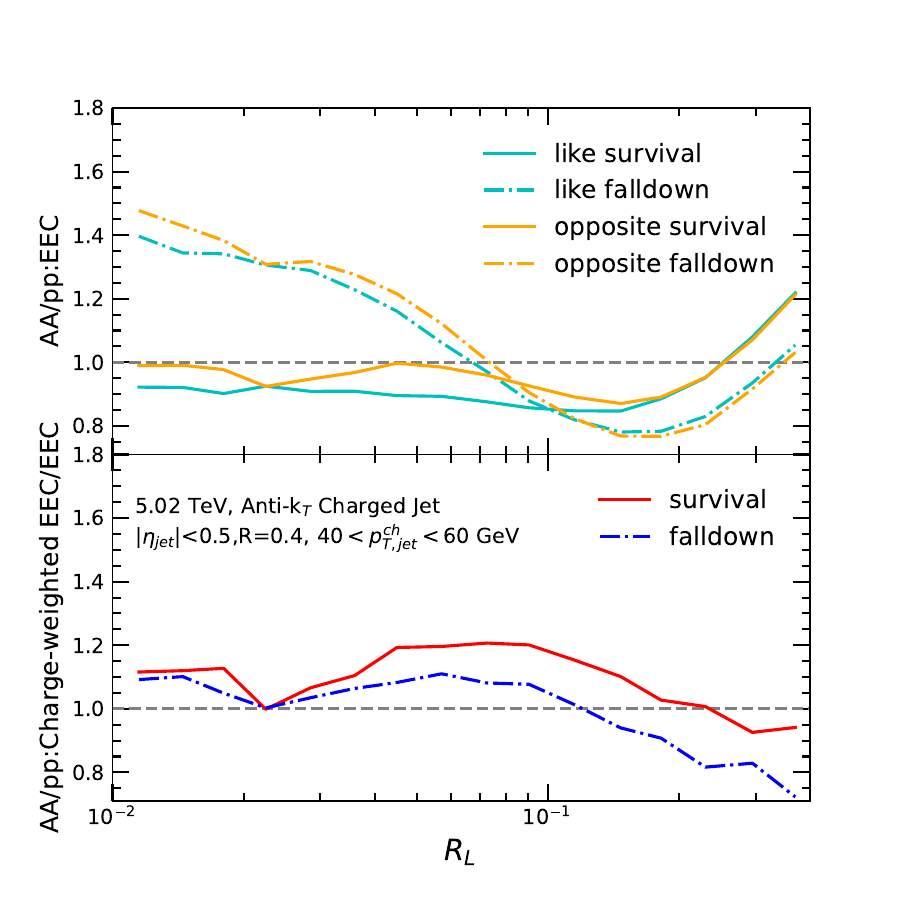}
\vspace{-1cm}
\caption{\textit{survival} and \textit{fall-down} contribution to the A+A/p+p ratios of like-charge and opposite-charge component of EEC (upper panel) and charge-weighted EEC/inclusive EEC (bottom panel) in central $(0-10\%)$ Pb+Pb collisions as functions of $R_\text{L}$ for  inclusive charged jets with a radius of $R=0.4$ in jet $p_{\rm T}$ interval $40-60$~GeV at $\sqrt{s_{NN}}=5.02$~TeV.}
\label{fig:slecbais}
\end{figure}

Next, one also needs to identify whether the plateau-like enhancement of charge correlations in the transition and small-$R_\text{L}$ region is related to the selection bias effect or not, since such an effect dominates the enhancement of the inclusive EEC in the small-$R_\text{L}$ region. Can it be canceled out with such a ratio? We then plot in the Fig.~\ref{fig:slecbais} (upper panel) the separate contributions of the \textit{survival} and \textit{fall-down} mechanisms to the A+A/p+p ratio for both like-charge and opposite-charge components, while the lower panel compares their impacts on the A+A/p+p ratio of the charge correlations. Here, \textit{survival} denotes jets in A+A collisions whose corresponding progenitor jets in p+p collisions fall within the same momentum interval, whereas \textit{fall-down} refers to A+A jets originating from p+p jets initially at higher momentum. In the decomposed view (upper panel), the effects of both mechanisms on the A+A/p+p ratio are consistent with our earlier EEC results—the \textit{fall-down} contribution enhances the distribution in the small-$R_\text{L}$ region. However, the lower panel reveals that the overall behaviors of the \textit{survival} and \textit{fall-down} cases are qualitatively similar, with only a minor difference in the small-$R_\text{L}$ region. At large $R_\text{L}$, the \textit{fall-down} contribution induces a more rapid decorrelations as $R_\text{L}$ increases. One can conclude that the plateau-like enhancement of the charge correlations in the transition and small-$R_\text{L}$ region is unrelated to selection bias, as this effect is canceled by constructing the ratio. 

\section{summary}\label{sec:4}

This work presents a systematic investigation of the medium modifications for the charge-dependent jet substructure using the charged-weighted EEC in p+p and $0-10\%$ central Pb+Pb collisions at $\sqrt{s_{\mathrm{NN}}} = 5.02$ TeV. We analyze charged hadron jets along with flavor-separated quark and gluon-initiated jets in the $40-60$~GeV momentum range with $R=0.4$. In p+p collisions, the charge correlations is consistently negative, reflecting the dominance of opposite-charge pairs due to charge conservation. In the transition and large $R_\text{L}$ regions, opposite-charge pairs exhibit progressively stronger decorrelations as $R_\text{L}$ increases. While Pythia simulations predict that opposite-charge correlations reaches its maximum in the small-$R_\text{L}$ region and remains constant, they fail to describe the experimentally observed decorrelations behavior as $R_\text{L}$ decreases. Meanwhile, a clear flavor dependence emerges: gluon-initiated jets exhibit less opposite-charge correlations in the transition and smaller $R_\text{L}$ region than quark-initiated jets, while more opposite-charge correlations in the larger $R_\text{L}$ region.

In Pb+Pb collisions, significant medium modifications are observed. Detailed calculation of the A+A/p+p ratio for the charged-weighted EEC reveals a modification pattern highly consistent with that of the inclusive EEC. A distinctive V-shaped modification appears exclusively in the opposite-charge component of the charge-weighted EEC, occurring at the same $R_\text{L}$ position regardless of jet flavor. Through factorization of the EEC into charged hadron pair multiplicity and average energy weighting distributions as functions of $R_\text{L}$, we identify that the enhanced energy weighting of opposite-charge pairs at small $R_\text{L}$ is where the observed V-shaped modification originates.

The A+A/p+p ratio for charge correlations shows a universal pattern across all jet types: plateau-like enhancement in the transition and small-$R_\text{L}$ regions, and increasing suppression with $R_\text{L}$ in the large-$R_\text{L}$ region. Jet quenching enhances opposite-charge correlations at small angles while reducing it at large angles, resulting in a steeper $R_\text{L}$ dependence, a rapid de-correlated behavior with the increasing $R_\text{L}$. The flavor-dependency seems to be canceled out when taking the ratio of charge-weighted distribution to the inclusive EEC. The distinctive V-shaped modification, which appeared exclusively in the opposite-charge component of the charge-weighted EEC, occurs at the same $R_\text{L}$ position regardless of jet flavor. The plateau-like enhancement of charge correlations in the transition and small-$R_\text{L}$ regions is unrelated to selection bias effects, based on the survival and fall-down contribution analysis. 

Our results establish charge-sensitive correlators as a powerful tool for probing jet quenching mechanisms of the charge correlations redistribution within jets in the quark-gluon plasma, and pave the way to further constrain the hadronization mechanism in the hot QCD medium.

{\bf Acknowledgments:}  This work is supported by a grant from the National Natural Science Foundation of China (Key Program) (No. 12535010)

\vspace*{-.6cm}

%


\end{document}